\documentclass[
reprint,
superscriptaddress,
amsmath,
amssymb,
aps,
prl,
floatfix,
]{revtex4-2}

\usepackage{graphicx}
\usepackage{dcolumn}
\usepackage{bm}
\usepackage{hyperref}
\usepackage{balance}
\usepackage{braket}
\usepackage{xcolor}
\usepackage[version=4]{mhchem}

\begin{document}

\title{Microscopic investigation of spin dynamics in the single-chain magnet Sr$_4$Mn$_2$CoO$_9$}

\author{G. Roy}
\email{gourabr22bs@rgipt.ac.in}
\affiliation{Rajiv Gandhi Institute of Petroleum Technology, Jais, Amethi 229304, Uttar Pradesh, India}

\author{S. Ghosh}
\affiliation{Rajiv Gandhi Institute of Petroleum Technology, Jais, Amethi 229304, Uttar Pradesh, India}

\author{M. Kumar}
\affiliation{Rajiv Gandhi Institute of Petroleum Technology, Jais, Amethi 229304, Uttar Pradesh, India}

\author{E. Kushwaha}
\affiliation{Rajiv Gandhi Institute of Petroleum Technology, Jais, Amethi 229304, Uttar Pradesh, India}

\author{J. Sannigrahi}
\affiliation{School of Physical Sciences, Indian Institute of Technology Goa, Ponda 403401, Goa, India}

\author{V. Caignaert}
\affiliation{Laboratoire CRISMAT, Université de Caen Normandie, ENSICAEN, CNRS UMR 6508, Normandie Univ., 14000 Caen, France}

\author{W. Prellier}
\affiliation{Laboratoire CRISMAT, Université de Caen Normandie, ENSICAEN, CNRS UMR 6508, Normandie Univ., 14000 Caen, France}

\author{D. T. Adroja}
\affiliation{ISIS Neutron and Muon Source, STFC, Rutherford Appleton Laboratory, Chilton, Oxon OX11 0QX, United Kingdom}
\affiliation{Highly Correlated Matter Research Group, Physics Department, University of Johannesburg, Auckland Park 2006, South Africa}

\author{D. Voneshen}
\affiliation{ISIS Neutron and Muon Source, STFC, Rutherford Appleton Laboratory, Chilton, Oxon OX11 0QX, United Kingdom}
\affiliation{Department of Physics, Royal Holloway, University of London, TW20 0EX, UK}

\author{V. Hardy}
\affiliation{Laboratoire CRISMAT, Université de Caen Normandie, ENSICAEN, CNRS UMR 6508, Normandie Univ., 14000 Caen, France}

\author{T. Basu}
\email{tathamay.basu@rgipt.ac.in}
\affiliation{Rajiv Gandhi Institute of Petroleum Technology, Jais, Amethi 229304, Uttar Pradesh, India}


\begin{abstract}

One-dimensional single-chain magnets offer a unique platform for studying the interplay of crystal-field effects, exchange interactions, and lattice dynamics. Here, we investigate spin excitations in Sr$_4$Mn$_2$CoO$_9$ using inelastic neutron scattering (INS) and theoretical modelling. INS reveals two low-energy magnetic excitations at $4$ and $7$ meV from Mn--Co--Mn spin chains, alongside higher-energy crystal-electric-field (CEF) excitations from two crystallographically inequivalent Co$^{2+}$ sites. Interestingly, these spin excitations persist at room temperature, demonstrating dynamic magnetic correlations in the absence of long-range order. Furthermore, the crystal-field modelling, based on Stevens operator formalism, reproduces well the CEF spectra, establishing Ising-like Kramers ground-state doublets with strong uniaxial magnetic anisotropy for both Co$^{2+}$ ions. In addition, the spin wave simulation using SpinW reproduces the spin excitation spectrum and reveals microscopic exchange interactions in two non-interacting Mn–Co–Mn spin chains. Finally, machine-learning lattice-dynamics calculations confirm the phonon spectrum and spin–phonon coupling. By projecting the exchange Hamiltonian onto CEF ground-state doublets, we estimate exchange-induced splittings matching the observed excitations. Thus, our results elucidate low-energy spin dynamics arising from combined crystal-field anisotropy and exchange interactions, with the persistent low-energy excitation providing a microscopic pathway for thermally activated spin relaxation. Furthermore, this work delivers a unified microscopic understanding of the interplay between crystal-field effects, magnetic exchange, and lattice dynamics in Sr$_4$Mn$_2$CoO$_9$, advancing insights into spin dynamics in low-dimensional transition-metal oxides.

\end{abstract}

 \maketitle

\section{Introduction}

\begin{figure}[!ht]
    \centering
    \includegraphics[width=0.45\textwidth]{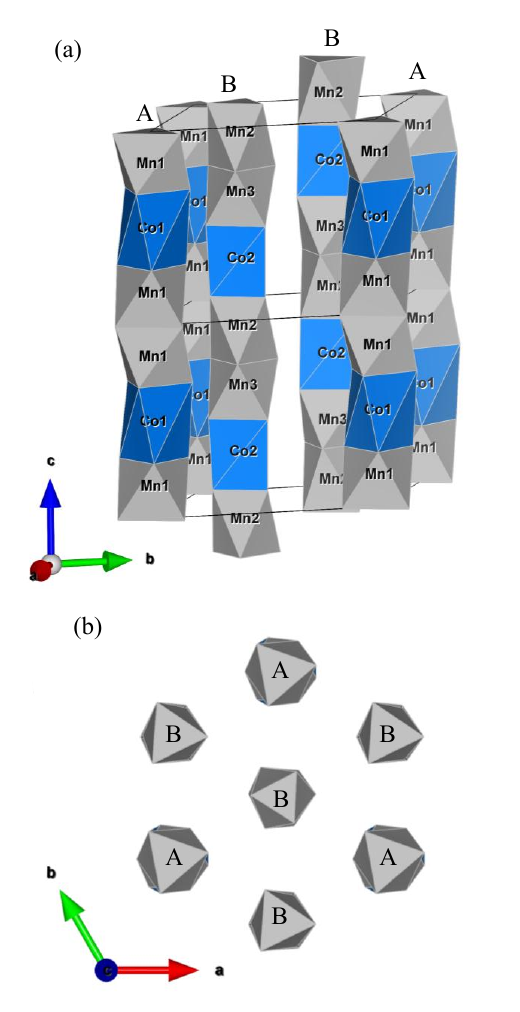}
    \caption{(a) Crystal structure of Sr$_4$Mn$_2$CoO$_9$ (Space group P$321$). The spin chains are formed by alternating \ce{MnO6} octahedra (grey polyhedra) and \ce{CoO6} trigonal prisms (cyan polyhedra). Sr atoms are omitted for clarity. The schematic illustrates the body-diagonal arrangement of the A and B spin chains within the unit cell. (b) View of the ab plane, showing the A and B spin chains arranged in a triangular lattice.}
    \label{fig:crystal_structure}
\end{figure}

The ability to control spin dynamics in low-dimensional strongly correlated electron systems has emerged as a prominent issue in modern condensed matter physics, given its fundamental importance and practical applications in information storage, spintronics, and quantum technology \cite{Bogani2008, Hymas2019, Najafi2019, MorenoPineda2021}. Systems with slow magnetic relaxation have received special interest because of their potential to store magnetic information over long time periods. Over more than the last three decades, single-molecule magnets (SMMs) have emerged as important quantum material that exhibits slow magnetisation relaxation below a characteristic blocking temperature, arising from high magnetic anisotropy and an energy barrier to spin reversal \cite{Vieru2024, Roy2026}. In these systems, the slow magnetic dynamics originate from individual molecules, where the combination of high spin states and strong spin–orbit coupling stabilises the magnetisation against thermal fluctuations. Later, the concept was expanded to lanthanide-based single-ion magnets (SIMs), in which slow magnetic relaxation results from an isolated magnetic ion with significant crystal-field-induced anisotropy, without requiring exchange-coupled magnetic clusters \cite{Ishikawa2003, Guo2018}. Although SMMs and SIMs have shown impressive slow magnetic relaxation at the molecular and single-ion levels, extending these capabilities to low-dimensional exchange-coupled systems has become a major research area, resulting in the development of single-chain magnets (SCMs) \cite{Ferbinteanu2005, Kishine2006}.

SCMs are one-dimensional magnetic systems with a slow magnetic relaxation resulting from the cooperative interplay of strong intrachain exchange interactions with significant uniaxial magnetic anisotropy.  Magnetisation reversal occurs via the nucleation and propagation of domain walls throughout the chain, rather than the independent reversal of individual spins \cite{PhysRevLett.95.237203}. Because the interchain interactions are weak enough, SCMs can exhibit slow magnetic relaxation in the absence of traditional three-dimensional long-range magnetic order. These distinguishing characteristics bring SCMs to the crossroads of molecule magnetism and extended magnetic solids, making them an important platform for investigating low-dimensional magnetism and next-generation magnetic information technology. Slow magnetic relaxation in SCMs can occur via a variety of microscopic mechanisms, including quantum tunnelling of magnetisation, direct, Raman, and Orbach processes \cite{Roy2026}. The primary relaxation pathway is determined by temperature, magnetic field, and spin-phonon coupling. At relatively high temperatures, the Orbach process is widely used to govern relaxation, in which thermally generated excited states help to reverse magnetisation. Determining the primary relaxation process thus offers important information about the magnetic anisotropy, exchange interactions, and spin dynamics of low-dimensional magnetic systems.

For Sr$_4$Mn$_2$CoO$_9$, the coexistence of strong antiferromagnetic intrachain exchange interactions, weak interchain coupling, and easy-axis anisotropy associated with trigonal-prismatic Co$^{2+}$ ions provides the required elements for SCM behaviour. Previous studies on the Sr$_{4-x}$Mn$_x$CoO$_9$ series revealed slow magnetic relaxation from bulk magnetic measurements \cite{Seikh2017, Seikh2018}, showing that Sr$_2$Ca$_2$Mn$_2$CoO$_9$ exhibits single-ion magnet (SIM) behaviour despite the presence of long-range magnetic order (LRO) \cite{Hardy2018, Basu2026}. However, Sr$_4$Mn$_2$CoO$_9$ does not exhibit LRO up to $1.7$ K \cite{Boulahya2003}; instead, it displays characteristic single-chain magnet (SCM) behaviour \cite{Seikh2017}. Interestingly, here the Sr/Ca ratio has a significant impact on adjusting metal-metal distances and magnetic exchange interactions, hence controlling the formation of long-range magnetic order in this series of materials. In Sr$_4$Mn$_2$CoO$_9$, trigonal-prismatic Co$^{2+}$ ions connect to octahedral Mn$^{4+}$ ions, generating two crystallographically different one-dimensional spin chains: chain A [Mn(1)--Co(1)--Mn(1)] and chain B [Mn(2)--Mn(3)--Co(2)] (Fig. ~1). The large Sr ions effectively divide the A and B chains, resulting in weak interchain magnetic interactions. This unique crystal structure provides an ideal platform for investigating how local crystal fields and exchange interactions cooperate to produce slow magnetic dynamics. Moreover, slow magnetic relaxation in SCMs has been extensively explored in coordination polymers and molecular systems over the last three decades, while its discovery in pure inorganic oxides is rare. 

The magnetic susceptibility does not exhibit a distinct anomaly associated with long-range magnetic ordering as in previous reports on Sr$_4$Mn$_2$CoO$_9$, where the one-dimensional nature of the spin chains and magnetic frustration suppress a clear bulk signature of the magnetic transition \cite{Boulahya2003, Seikh2017}. Moreover, the Curie--Weiss fit from bulk magnetization yields an effective paramagnetic moment of $\mu_{\rm eff}=7.1~\mu_{\rm B}$ per formula unit, which is similar to the expected value for two Mn$^{4+}$ ions ($\mu_{\rm eff}=3.87~\mu_{\rm B}$ per Mn) and one high-spin Co$^{2+}$ ion ($\mu_{\rm eff}=4.80~\mu_{\rm B}$ per formula unit). The higher magnetic moment of Co$^{2+}$ compared to its spin-only value ($\mu_{\rm eff}=3.87~\mu_{\rm B}$) indicates a partially unquenched orbital contribution. The Curie-Weiss fit results in a Weiss temperature of $\Theta_{\rm CW}=-91$~K, indicating antiferromagnetic magnetic interactions \cite{Boulahya2003}. Thus, bulk magnetic measurements provide essential information on magnetic ordering, magnetic anisotropy, and relaxation dynamics, but it does not directly expose the microscopic quasiparticle excitations that cause such properties. In contrast, inelastic neutron scattering (INS) is one of the few experimental techniques capable of simultaneously probing spin-wave excitations and crystal electric field (CEF) levels over a wide energy range. Consequently, INS provides a scope for microscopic understanding of the origin of slow magnetic relaxation in low-dimensional magnets \cite{Basu2026, Waldmann2007, Pieper2010}. This approach allows for the determination of one-dimensional magnetic correlations, intrachain exchange interactions, magnetic anisotropy, and spin-phonon coupling, which collectively govern the slow magnetic relaxation in this unique inorganic single-chain magnet.

\section{Experimental Details}

The polycrystalline Sr$_4$Mn$_2$CoO$_9$ sample used in the present work was synthesised by the conventional solid-state reaction method by Vincent Hardy's group, following the procedure reported in Ref.~\cite{Seikh2017, Seikh2018}. The same single-phase polycrystalline sample, previously employed for the exploration of the Sr$_{4-x}$Ca$_x$Mn$_2$CoO$_9$ series \cite{Seikh2017, Seikh2018}, was used for the present inelastic neutron scattering (INS) measurements. Fig. 1 depicts the crystal structure of Sr$_4$Mn$_2$CoO$_9$, visualised with the VESTA software \cite{Momma2011}. Chains A and B are clearly distinguished.

INS measurements were conducted on the MERLIN time-of-flight spectrometer at ISIS, UK, with a $9$ g polycrystalline sample mounted in an annular geometry in a cylindrical aluminium container with a diameter of $ 30$ mm \cite{Caignaert2018_RB1820225}. Measurements were performed using MERLIN in multi-$E_i$ mode, with incident neutron energies of $E_i=82.1$, 30, 15.4, and 9.4~meV, using a Gd-Fermi chopper operating at 400~Hz. Data were taken from 7.6--300 K. To investigate low-temperature spin excitations, simultaneous INS measurements were carried out on the LET cold-neutron time-of-flight spectrometer at ISIS \cite{Caignaert2018_LET}. The $9$ g polycrystalline sample, mounted in an aluminium can, was measured using LET in the multi-$E_i$ mode with incident neutron energies of $E_i=12$, $5.1$, $2.8$, and $1.8$ meV, employing the High Flux (HF) chopper configuration ($240/120$).  The INS data were analysed using the Mantid and DAVE software packages \cite{Arnold2014, Azuah2009}.

The magnetic excitation spectrum was modelled using the modelled based on the linear spin wave theory (LSWT) using the SpinW \cite{Toth2015}, a MATLAB-based library that calculates and visualises spin-wave dispersions within the framework of linear spin-wave theory.

The McPhase package was used to calculate the crystal-field (CEF) of high-spin Co$^{2+}$ ions occupying trigonal-prismatic sites \cite{Rotter2012}. The Stevens operator formalism was used to generate the CEF Hamiltonian, and the crystal-field parameters were optimised to reproduce the low-energy INS excitations seen in the experiment. The resulting energy-level scheme and eigenfunctions were used to explain the observed crystal-field excitations.

Phonon calculations were carried out using the INSPIRED software package, which makes use of pre-trained machine-learning force fields (MLFFs) created within the MatterSim framework \cite{Yang2024}. The improved crystal structure and momentum-dependent phonon spectra were then calculated and compared directly to the INS results \cite{Roy2026_BDRO}.

\section{Results and Discussions}

\subsection{Inelastic Neutron Scattering}

Inelastic neutron scattering (INS) is an important technique for investigating spin dynamics, low-energy excitations, phonon excitation, the crystal electric field (CEF), and understanding the nature of magnetic exchange interactions. The INS spectra with an incident energy ($E_i = 12$~meV) were recorded in the LET instrument at temperatures of 2~K, 5~K, 7~K, 10~K, and 12.5~K to understand the low-energy magnetic excitation. Figs. ~2(a)--(b) represent the contour colour plots of these spectra at $2$ K and $12.5$ K, revealing distinct excitations around $4$ and $7$ meV in the low-momentum transfer ($Q$) regions $1.4$-$3$ A$^{-1}$. The corresponding Intensity (I) vs Energy Transfer(E) $1$D plot is present in Fig. ~2(c). The intensity of these low-energy excitations decreases with increasing Q \cite{Roy2026SM}, as the magnetic form factor decreases exponentially with increasing Q (see Fig. S1 in the Supporting Information (S.I.)) . This confirms the excitations originate from magnetic excitation \cite{Roy2026_BDRO}. The broad magnetic excitation with a big spin gap comes from dispersive spin wave excitation. In Fig. ~2, the $4$ meV excitation is clearly visible and diminishes with incresing temperature. Here in Sr$_4$Mn$_2$CoO$_9$, even in the absence of long-range order, the interchain interaction is not present, but the presence of intrachain antiferromagnetic interaction give rise to spin excitation. The persistence of the broad magnetic excitation over the measured temperature range indicates that the dominant antiferromagnetic intrachain exchange interactions remain robust. The broad spectral response arises from powder-averaged dispersive collective spin excitations associated with these strong one-dimensional magnetic correlations, while the weak interchain coupling prevents the establishment of long-range magnetic order. The corresponding Bose-corrected INS spectra, analogous to those shown in Figs.~2(a)--(c), are presented in Figs.~S2(a)--(c) of the S.I. The Bose correction accounts for the temperature-dependent thermal population factor, enabling a more direct comparison of the intrinsic magnetic response at different temperatures.

\begin{figure*}[t]
    \centering
    \includegraphics[width=1\textwidth]{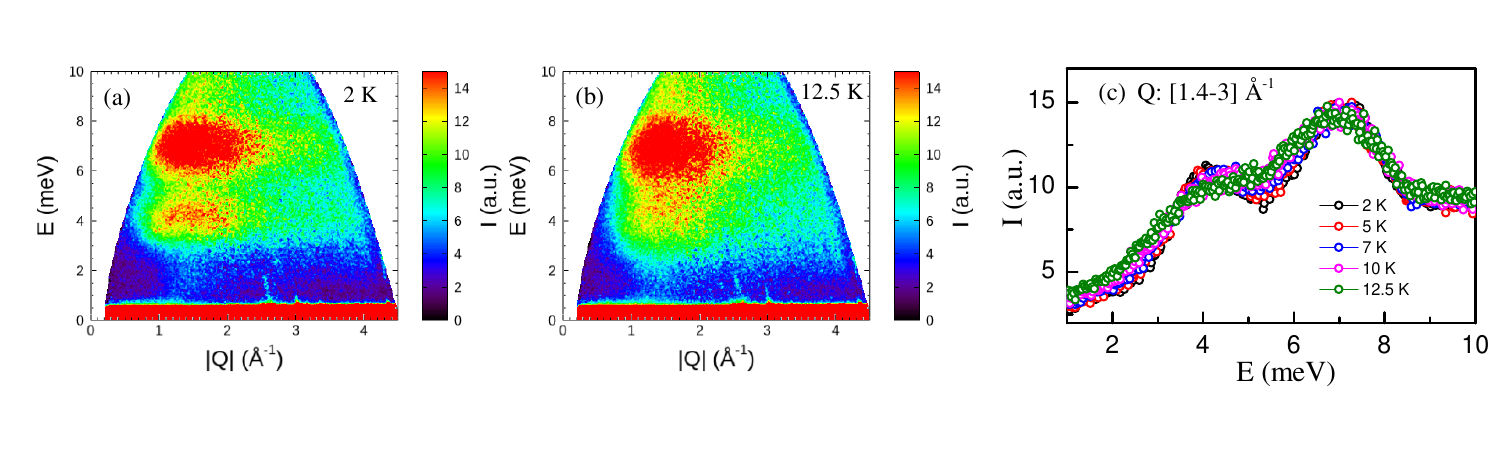}
    \caption{(a,b) Colour contour maps of the inelastic neutron scattering (INS) intensity measured using the LET spectrometer with an incident energy of $E_i = 12$ meV at 2 K and 12.5 K, respectively. (c) Energy dependence of the INS intensity integrated over the $Q$ range of $1.4$--$3.0~\mathrm{\AA^{-1}}$ at various temperatures.}

    \label{LET_Magnon}
\end{figure*}

To further investigate the magnetic excitations and their interplay with phonon modes, INS measurements were performed on the MERLIN spectrometer with an incident energy of $E_i = 30$~meV at temperatures of 7~K, 10~K, 18~K, 28~K, and 300~K. Figs. ~3(a)--3(c) present the corresponding colour contour maps recorded at 7~K, 10~K, and 28~K, respectively. The spectra reveal magnetic excitations in the low-momentum transfer ($Q$) region ($1$--$4$~\AA$^{-1}$) (where I decay with Q as shown in Fig. S3 from S.I), extending over an energy range of approximately $5$--$13$~meV, which is a dispersive-like spin excitation. The corresponding one-dimensional energy cuts [Fig.~3(d)] exhibit peaks centred around $7$ and $11$~meV. Compared with the low-energy INS measurements ($E_i=12$~meV), where a magnetic excitation near $4$~meV is clearly resolved, this feature is not distinctly observed in the higher incident-energy data because of the reduced energy resolution. However, an additional weak excitation near $11$~meV becomes evident. These observations suggest the presence of three magnetic excitation modes centred around $4$, $7$, and $11$~meV. Near $16-17$ meV, there is also weak magnetic excitations at low-T, but at high-T it is dominated by the phonon. With increasing temperature, the magnetic intensity gradually decreases; nevertheless, the spin excitation remains clearly detectable even at 300~K. The persistence of these magnetic excitations to room temperature demonstrates the robustness of the intrachain antiferromagnetic exchange interactions and magnetic anisotropy, indicating that dynamic spin correlations survive over a wide temperature range despite the absence of long-range magnetic order.

\begin{figure*}[t]

    \centering
    \includegraphics[width=1\textwidth]{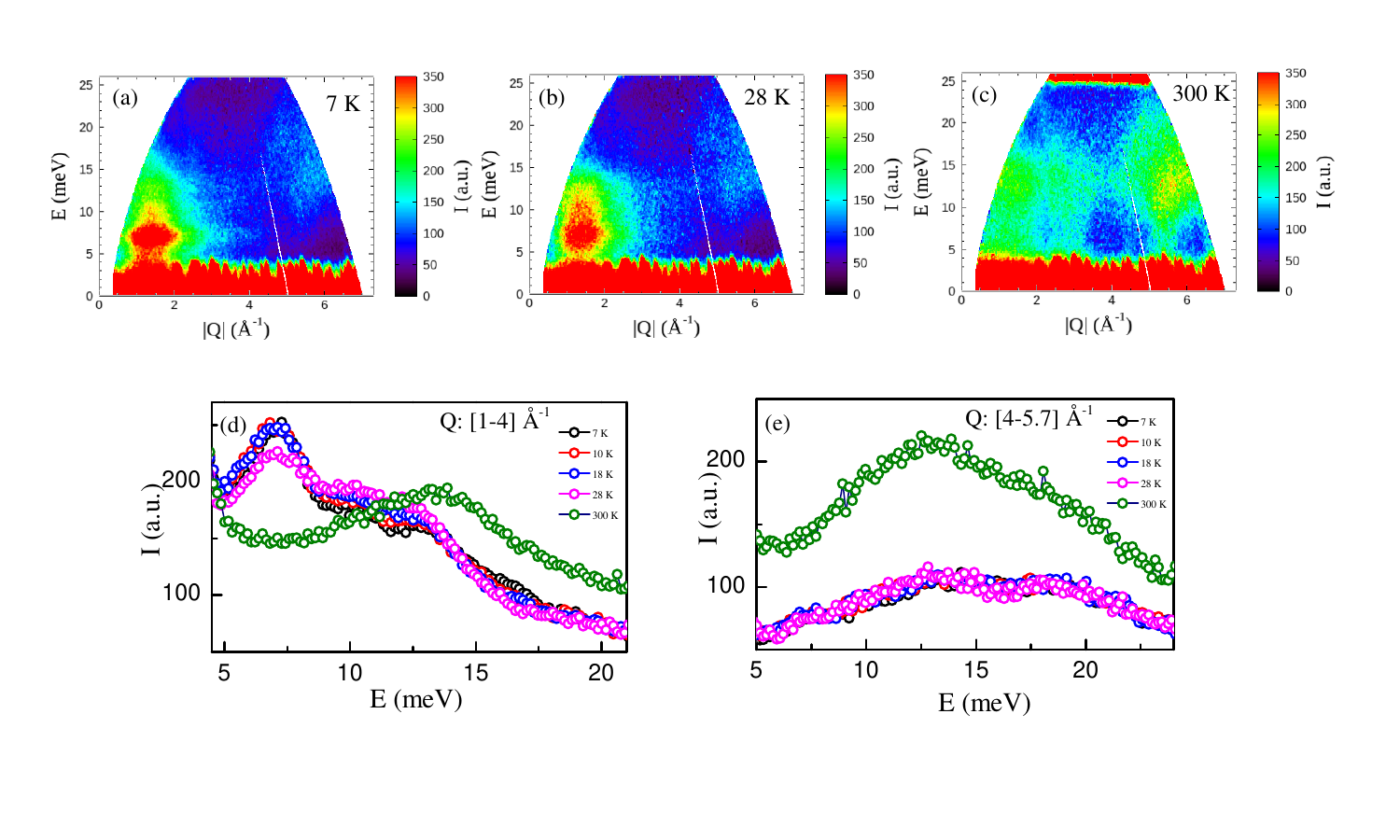}
    \caption{(a--c) Colour contour maps of the inelastic neutron scattering (INS) intensity measured using the MERLIN spectrometer with an incident energy of $E_i = 30$ meV at 7 K, 28 K, and 300 K, respectively. (d,e) INS intensity as a function of energy transfer, integrated over the $Q$ ranges of $1$--$4$ and $4$--$5.7~\mathrm{\AA^{-1}}$, respectively, at various temperatures.}
    
    \label{MERLIN_Magnon}
\end{figure*}

\begin{figure*}[t]
    \centering
    \includegraphics[width=1\textwidth]{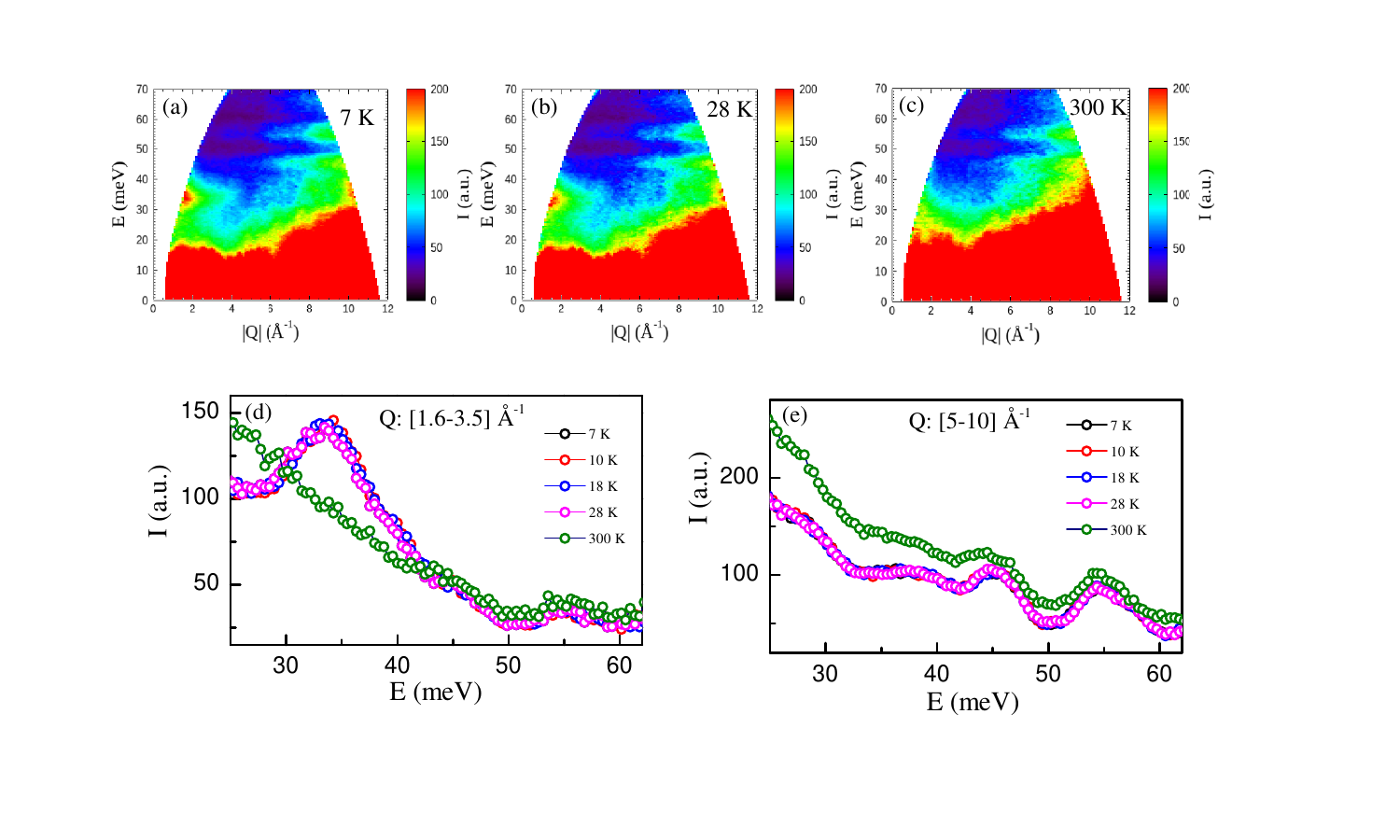}
    \caption{(a--c) Colour contour maps of the inelastic neutron scattering (INS) intensity measured using the MERLIN spectrometer with using an incident energy of $E_i = 82.1$ meV at 7 K, 28 K, and 300 K, respectively. (d,e) INS intensity as a function of energy transfer, integrated over the $Q$ ranges of $1.6$--$3.5$ and $5$--$10~\mathrm{\AA^{-1}}$, respectively, at various temperatures.}
    \label{MERLIN_CEF}
\end{figure*}

In contrast, phonon excitations dominate in the high-$Q$ region following Inensity (I) proportional to Q$^2$ \cite{Roy2026SM}, and strengthen with incresing temperature due to enhanced thermal lattice vibrations (see Fig. S4 from S.I) \cite{Roy2026_BDRO}. A careful look at the contour plot of INS spectra (Figs. ~3(a)--(c)) in the high-$Q$ region also exhibits excitations around the same energy region, which become more prominent with increasing temperature. This is further documented in the one-dimensional cut over a $Q$ ranging from 4--5.7~\AA$^{-1}$ (Fig. ~3(e)). This high-$Q$ feature is indicative of phonon excitation around $7-21$meV. Interestingly, the magnon peak shifts to higher energy with increasing temperature (see Fig. ~3(d)), while phonon modes are clearly present in the same energy window (Fig. ~3(e)). At 300~K, magnon and phonon excitations coexist in almost the same energy window. This anomalous upward shift of the magnon mode toward the phonon energy scale with temperature is possibly due to magnon--phonon coupling. The migration of magnon excitations to higher energies with increasing temperature signifies a temperature-induced renormalisation of the magnetic Hamiltonian. Moreover, the low-energy magnetic modes centred around $4$ and $7$~meV exhibit a much stronger temperature dependence, with their intensities decreasing significantly at higher temperatures [Fig.~3(d)], in the context of spin excitation. In contrast, the $11$~meV excitation does not show a comparable reduction in intensity. This behaviour is likely due to an increasing phonon contribution in the same energy range, as evidenced by the enhanced high-$Q$ phonon scattering at elevated temperatures. As the temperature increases, the phonon spectral weight progressively overlaps with the magnetic excitation around $11$~meV, leading to a mixed spectral response. Consequently, while the magnetic contribution decreases with temperature, the increasing phonon intensity must effect in spin intensity, causing the $11$~meV feature to remain comparatively unchanged. This coexistence and increasing overlap of magnetic and lattice excitations at higher temperatures further suggest a strong interplay between spin and phonon degrees of freedom. The corresponding Bose-corrected INS spectra, analogous to those shown in Figs.~3(a)--(e), are presented in Figs.~S5(a)--(e) of the S.I.

To probe the higher-energy magnetic excitations, INS measurements were performed on the MERLIN spectrometer with an incident energy of $E_i = 82.1$~meV. Figs. ~4(a)--4(c) show the corresponding colour contour maps recorded at 7~K, 28~K, and 300~K, respectively, while the corresponding one-dimensional energy cuts are presented in Fig.~4(d). A broad excitation is observed in the low-$Q$ region ($Q \approx 1.6$--$3.5$~\AA$^{-1}$) , whose intensity decreases with increasing momentum transfer (see Fig. S6 from S.I), confirming its magnetic origin\cite{Roy2026SM}. In contrast, the excitation appears nearly dispersionless in the colour maps, which is a characteristic signature of localised crystal electric field (CEF) excitations arising from single-ion anisotropy. The one-dimensional INS spectra reveal two well-defined excitation modes centred at approximately $33$ and $55$~meV. These CEF excitations originate from the spin--orbit-coupled crystal-field splitting of the $\mathrm{Co}^{2+}$ ions occupying the trigonal prismatic Co(1) sites (point group $D_{3}$) and Co(2) sites (point group $C_{3}$). Similar CEF excitations have also been reported in other Co-based magnetic systems \cite{Basu2026, Yuan2020, Ringler2022}. A detailed theoretical analysis of the CEF level scheme and spectral splitting, based on crystal-field modelling, is presented in the following section. In the high-$Q$ region ($Q \approx 5$--$10$~\AA$^{-1}$), where the scattering intensity increases approximately as $Q^{2}$ (see Fig. S7 from S.I), strong phonon excitations are observed around $28$, $36$, $44$, and $54$~meV (see Fig. 4(e)). Notably, the excitation near $54$--$55$~meV is present in both the low- and high-$Q$ regions, indicating an overlap between the magnetic CEF excitation and the phonon mode. The coexistence of these excitations in the same energy window suggests a possible coupling between the crystal-field excitations and lattice vibrations (vibronic or CEF--phonon coupling). Further evidence for this interaction is discussed in the subsequent theoretical analysis.

\subsection{Spin-Wave Simulations}

The spin-wave calculations were performed using the SpinW package \cite{Toth2015}, which implements linear spin-wave theory for magnetic systems. The crystallographic structure of Sr$_4$Mn$_2$CoO$_9$ was constructed using the experimental lattice parameters and space group ($P321$). The magnetic lattice consists of five inequivalent magnetic ions, namely Mn1, Mn2, Mn3, Co1, and Co2. The magnetic Hamiltonian consists of seven isotropic Heisenberg exchange interactions ($J_1$--$J_7$) plus a uniaxial single-ion anisotropy term ($D=-2.25$~meV), assigned to each magnetic ion. It is represented as

\begin{equation}
\mathcal{H}
=\sum_{\langle i,j\rangle} J_{ij}\,\mathbf{S}_i\cdot\mathbf{S}_j
+\sum_i D(S_i^{z})^{2},
\end{equation}

where $J_{ij}$ denotes the exchange interactions between neighbouring magnetic ions. The dynamical structure factor $S(Q,\omega)$ was computed within linear spin-wave theory and powder averaged using Monte Carlo sampling. Instrumental energy resolution, magnetic form factor, and powder averaging were incorporated to enable direct comparison with the experimental inelastic neutron scattering spectra. From the simulations, the extracted $J$ and $D$ values are listed in Table~I. The simulated spin-dispersion curves and the corresponding contour plot of the spin-excitation spectra are shown in Figs.~5(a) and 5(b), respectively, which reveal clear excitations around $4.8$, $8$, and $10.3$~meV up to $Q = 4~\text{\AA}^{-1}$. These simulated spin excitations match well with the experimental results, as shown in Figs.~2 and 3, where magnetic excitations are observed around $4$, $7$, and a weak $11$~meV.

The anisotropy $D$ in the $3 \times 3$ anisotropy tensor is found to be

\[
D =
\begin{pmatrix}
0 & 0 & 0 \\
0 & 0 & 0 \\
0 & 0 & -2.25
\end{pmatrix}.
\]

From the SpinW calculations, we obtained $D_{aa}=D_{bb}=0$ and $D_{cc}\approx -2.25$~meV, indicating that the spin environment possesses axial symmetry with easy-axis uniaxial anisotropy along the crystallographic $c$-direction. The negative value of $D_{cc}$ corresponds to easy-axis anisotropy, stabilising Ising-like spin alignment and opening a finite spin gap in the magnetic excitation spectrum, thereby revealing an Ising spin-chain configuration.

The exchange interactions calculated from the SpinW simulations are summarised in Table~I. All calculated exchange interactions are antiferromagnetic, with $J_{\mathrm{Mn-Mn}} > J_{\mathrm{Co-Mn}}$.

\begin{table}[ht]
\centering
\caption{Exchange interactions obtained from SpinW simulations for chain-A and chain-B, where nn is the nearest neighbour and nnn is the next-nearest neighbour.}
\begin{tabular}{l@{\hspace{0.3cm}}l@{\hspace{0.3cm}}c}
\hline
\textbf{Chain} & \textbf{Metal--Metal Interaction} & \textbf{Exchange $J$ (meV)} \\
\hline
B & $J_1$ [nn] Mn(2)--Mn(3) & 1.60 \\
A & $J_2$ [nn] Mn(1)--Mn(1) & 1.40 \\
A & $J_3$ [nn] Mn(1)--Co(1) & 1.10 \\
B & $J_4$ [nn] Mn(3)--Co(2) & 0.14 \\
B & $J_5$ [nn] Mn(2)--Co(2) & 0.80 \\
B & $J_6$ [nnn] Mn(2)--Co(2) & 0.40 \\
B & $J_7$ [nnn] Mn(3)--Co(2) & 0.20 \\
\hline
\end{tabular}
\end{table}

\begin{figure}[!ht]
    \centering
    \includegraphics[width=.6\textwidth]{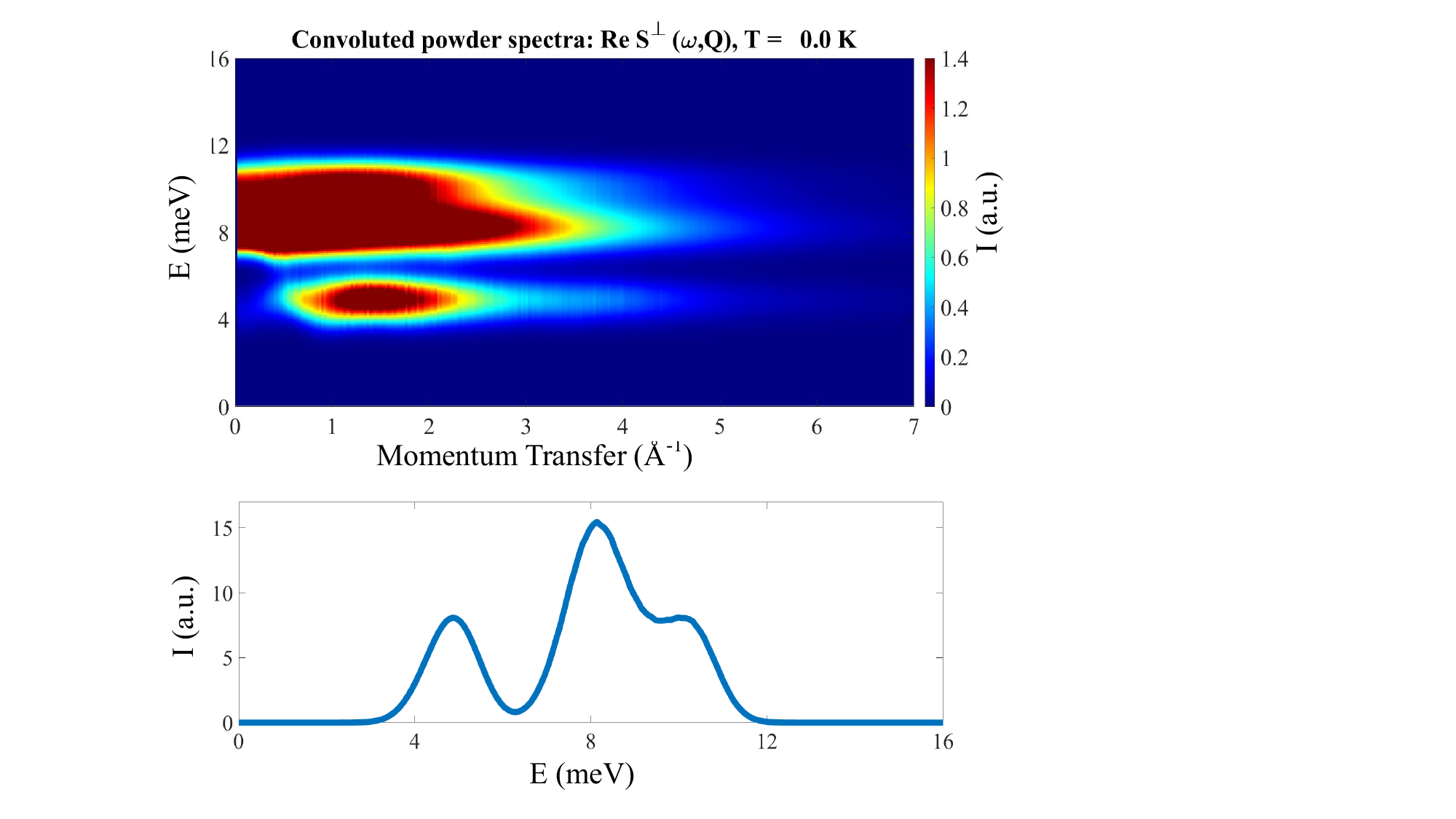} 
    \caption{ (a) Colour contour map of the simulated powder-averaged spin-wave intensity as a function of momentum transfer ($Q$) and energy transfer, calculated using SpinW. (b) Simulated spin-wave intensity as a function of energy transfer, integrated over the low-$Q$ region.
}
    \label{fig:SpinW}
\end{figure}

\subsection{Machine-Learning Modelling of Phonon Excitations}

We have theoretically calculated the phonons across the full $|Q|$ range to separate the phonon contributions from the magnetic signals observed in the INS data using a machine-learning-based force-field calculation \cite{Yang2024}. For this, we used the crystal structure of Sr$_4$Mn$_2$CoO$_9$ and optimised the structure. Fig. ~6(a) shows the calculated phonon spectrum, which extends up to 25~THz and exhibits no imaginary frequencies, confirming the dynamical stability of the optimised structure. After obtaining the optimised structure, we calculated the powder-averaged neutron scattering intensity convolved with the instrumental energy and momentum resolutions. This calculation also included multiphonon processes and the temperature-dependent thermal population of phonons. The simulated intensity [Fig.~6(b)] with $E_i = 30$~meV shows that phonon scattering dominates the $|Q|$ region from $3$ to $6.5~\text{\AA}^{-1}$ in the energy range of $7$--$24$~meV, which closely matches the experimental phonon excitations shown in Fig. ~3(e). Further, the simulated spectrum with $E_i = 80$~meV [Fig. ~6(c) and 6(d)] allows us to identify well-separated phonon excitations in the high-$Q$ region ($4$--$11~\text{\AA}^{-1}$) at approximately $25$--$29$, $33$, $35$--$38$, $40$--$45$, and $51$--$57$~meV. These calculated phonon excitations almost match the experimental excitations as shown in Fig. ~4(e). The machine-learning-based phonon calculations further validate the experimental observations and enable a clear distinction between phonon excitations and magnetic excitations, such as spin excitations (Fig.~3) and crystal electric field (CEF) excitations (Fig.~4).

\begin{figure*}[t]
    \centering
    \includegraphics[width=1\textwidth]{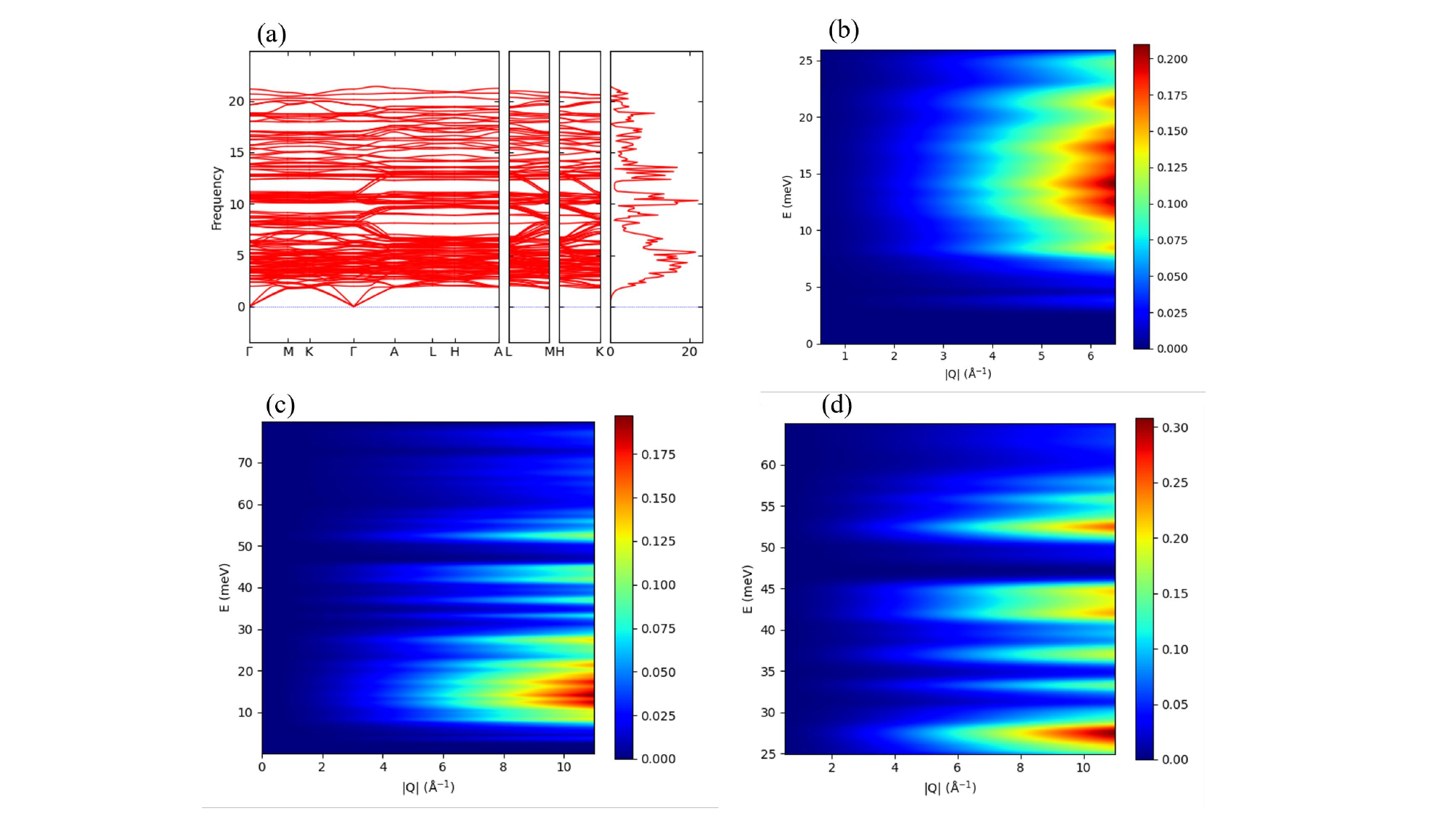}
    \caption{(a) Calculated phonon dispersion of Sr$_4$Mn$_2$CoO$_9$ obtained using a pre-trained machine-learning force field (MLFF). The vertical axis represents the phonon energy (THz), and the horizontal axis follows the standard high-symmetry directions in reciprocal space. (b,c) Simulated inelastic neutron scattering (INS) phonon intensity calculated using the MLFF for incident energies of $E_i=30$ and $82.1$ meV, respectively. (d) Enlarged view of the high-energy phonon excitations in panel (c) over the energy range of 25--65 meV.}
    \label{fig:MLFF}
\end{figure*}

\subsection{Crystal Field Calculations Using McPhase}

To establish the microscopic origin of the crystal-electric-field (CEF) excitations of the Co$^{2+}$ ions in Sr$_4$Mn$_2$CoO$_9$, single-ion calculations were performed using the \textsc{McPhase} package \cite{Rotter2012}. The Co$^{2+}$ ion has a $3d^7$ electronic structure. The free-ion Hamiltonian is defined by taking into account intraatomic electron-electron Coulomb interactions as well as spin-orbit coupling. The Coulomb interaction is represented by the Slater integrals $F^{2}=10563.1$ meV and $F^{4}=6820.57$ meV, which specify the free-ion multiplet structure. The coupling constant $\zeta = 66.21$ meV indicates the spin-orbit interaction. These free-ion interactions develop the Russell--Saunders multiplets before the crystal-field potential generated by the oxygen ligands is applied. The radial expectation values $\langle r^{2}\rangle =1.2855$ and $\langle r^{4}\rangle =3.9526$ describe the spatial extent of the $3d$ electron density and are used to analyze the electrostatic interaction between the Co$^{2+}$ ion and the surrounding ligand charges. The local trigonal-prismatic crystal field then lifts the free-ion multiplets' degeneracy, resulting in several energy levels separated by the crystal field. The electronic structure depends on intra-atomic Coulomb interactions, spin-orbit coupling, and the electrostatic crystal field.

The total single-ion Hamiltonian is expressed as,

\begin{equation}
\hat{H}
=
\hat{H}_{\rm Coulomb}
+
\hat{H}_{\rm SO}
+
\hat{H}_{\rm CF},
\label{eq:Htotal}
\end{equation}

where $\hat{H}_{\rm Coulomb}$ describes the intra-atomic electron-electron repulsion, $\hat{H}_{\rm SO}$ represents the spin-orbit interaction,

\begin{equation}
\hat{H}_{\rm SO}
=
\zeta\,\mathbf{L}\cdot\mathbf{S},
\label{eq:Hso}
\end{equation}

and $\hat{H}_{\rm CF}$ denotes the crystal-field Hamiltonian. In Stevens operator notation,

\begin{equation}
\hat{H}_{\rm CF}
=
\sum_{k,q}
B_k^q
O_k^q,
\label{eq:Hcf1}
\end{equation}

\begin{figure*}[t]
    \centering
    \includegraphics[width=1\textwidth]{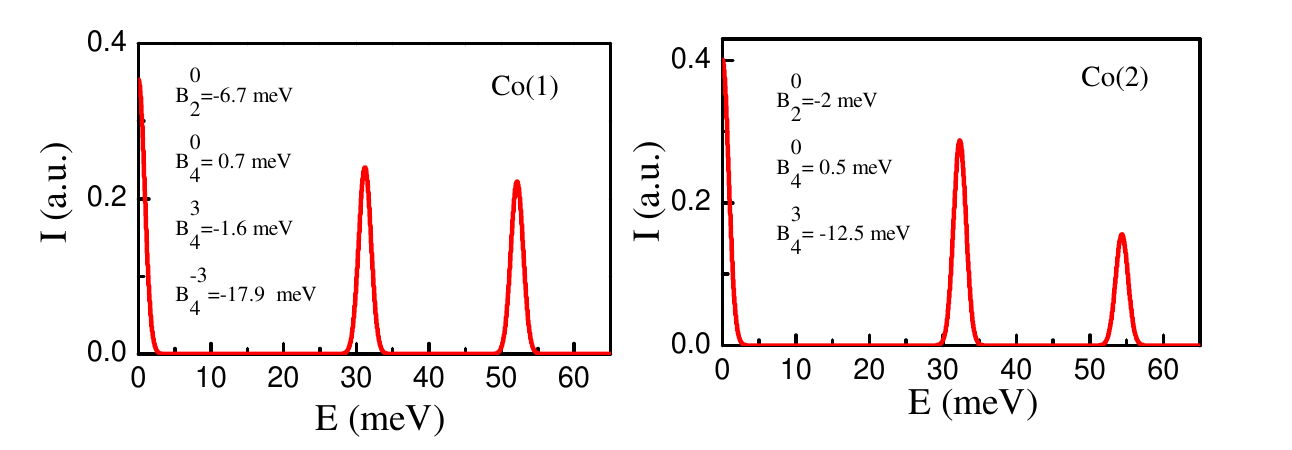}
    \caption{Calculated crystal electric field (CEF) excitation spectra of (a) Co(1) and (b) Co(2) in Sr$_4$Mn$_2$CoO$_9$ at $300$K, obtained using the McPhase program with the optimized Stevens crystal-field parameters determined from the CEF analysis.}
    \label{fig:McPhase}
\end{figure*}

Fig. 7 summarizes the optimised crystal-field parameters for Co(1) (point group D$3$) and Co(2) (point group C$3$) crystallographic sites after refinement. Hund's rules for the free Co$^{2+}$ ($3d^7$) ion yield the ground term $^4F$ with orbital angular momentum $L=3$ and spin $S=3/2$, resulting in the free-ion multiplets $J=9/2$, $7/2$, $5/2$, and $3/2$. In the present intermediate-coupling technique, the Hamiltonian is generated in the coupled basis $\left|L, S, J, M_J\right\rangle$, giving a convenient description of the spin-orbit interaction. Because the crystal-field interaction does not commute with $\hat{J}^{2}$, distinct $J$ multiplets are mixed, and $J$ is no longer a good quantum number for the final crystal-field eigenstates. As a result, the eigenfunctions are written as linear combinations of various $\left|L, S, J, M_J\right\rangle$ basis states \cite{Abragam2012, Newman2000}. The Hamiltonian matrix is then diagonalised by solving

\begin{equation}
\hat{H}
|\psi_n\rangle
=
E_n
|\psi_n\rangle,
\label{eq:sch}
\end{equation}

yielding the eigenvalues $E_n$ and eigenfunctions $|\psi_n\rangle$. For the $3d^7$ configuration, the intermediate-coupling calculation generates a total of $28$ crystal-field states corresponding to the $(2S+1)(2L+1)=4\times7=28$ basis functions. The calculated CEF energies for the two crystallographically inequivalent Co sites are shown in Fig.~7. For Co(1), the first three Kramers doublets occur at approximately $0$, $31.1$, and $52.1$ meV, whereas for Co(2) they are located at $0$, $32.3$, and $54.3$ meV (Fig. $7$). 

The calculated ground state wavefunction of Co(1) is,

\begin{align}
|\psi_0\rangle
=&
0.513
|^4F_{9/2},7/2\rangle
+0.206
|^4F_{9/2},1/2\rangle
\nonumber\\
&
+0.126
|^4F_{9/2},-5/2\rangle
+0.098
|^4F_{7/2},-5/2\rangle
+\cdots .
\end{align}

with the corresponding time-reversed partner

\begin{align}
|\psi_0'\rangle
=&
0.513
|^4F_{9/2},-7/2\rangle
+0.206
|^4F_{9/2},-1/2\rangle
\nonumber\\
&
+0.126
|^4F_{9/2},5/2\rangle
+0.098
|^4F_{7/2},5/2\rangle
+\cdots .
\end{align}

The dominant contributions to the Co(1) ground-state wavefunction arise from the
$|^4F_{9/2},7/2\rangle$, $|^4F_{9/2},1/2\rangle$, $|^4F_{9/2},-5/2\rangle$, and
$|^4F_{7/2},-5/2\rangle$ basis states, indicating substantial mixing between the
$J = 9/2$ and $J = 7/2$ multiplets due to the comparable strengths of the
spin--orbit and crystal-field interactions.

Similarly, the ground-state Kramers doublet of Co(2) is predominantly composed of

\begin{align}
|\psi_0\rangle
=&
-0.817
|^4F_{9/2},7/2\rangle
-0.348
|^4F_{9/2},1/2\rangle
\nonumber\\
&
+0.281
|^4F_{9/2},-5/2\rangle
-0.245
|^4F_{7/2},7/2\rangle
+\cdots .
\end{align}

together with its time-reversed partner

\begin{align}
|\psi_0'\rangle
=&
-0.817
|^4F_{9/2},-7/2\rangle
+0.348
|^4F_{9/2},-1/2\rangle
\nonumber\\
&
+0.281
|^4F_{9/2},5/2\rangle
+0.245
|^4F_{7/2},-7/2\rangle
+\cdots .
\end{align}

The computed crystal-field excitation energies for both Co(1) and Co(2) accurately approximate the experimentally observed inelastic neutron-scattering peaks at $32$ and $54$ meV. This demonstrates that the optimised crystal-field parameters precisely reflect the low-energy electronic excitations of Co$^{2+}$. The estimated eigenfunctions show significant mixing between the $^4F_{9/2}$, $^4F_{7/2}$, $^4F_{5/2}$, and higher multiplets, especially for excited states over $\sim100$ meV. Multiplet mixing is characteristic of Co$^{2+}$ ions bearing significant spin-orbit coupling in a low-symmetry ligand environment \cite{Waldmann2026}, which can play an important role in generating the strong magnetic anisotropy found in Sr$_4$Mn$_2$CoO$_9$. The resulting crystal-field level scheme provides a microscopic explanation of the experimentally observed CEF excitations and serves as the basis for understanding the compound's magnetic anisotropy and low-temperature magnetic properties.

\subsection{Exchange-Induced Spin Gap and Relaxation Mechanism}

In the previous section, we examined the combined effects of spin--orbit coupling (SOC) and the crystal electric field (CEF) to determine the ground-state wavefunctions of Co(1) (from chain A: Mn1-Co1-Mn1) and Co(2) (from chain B: Mn2-Mn3-Co2). Based on the results, we apply the exchange and anisotropy Hamiltonians to the respective ground states of Co(1) and Co(2). As a result, we can anticipate how the possible spin gap will be formed in chains A and B. In general, the microscopic Hamiltonian of the Mn--Co--Mn chain is expressed as,

\begin{equation}
H=
J_{12}\mathbf{S}_1\cdot\mathbf{S}_2
+
J_{13}\mathbf{S}_1\cdot\mathbf{J}
+
J_{23}\mathbf{S}_2\cdot\mathbf{J}
+
DJ_z^2,
\label{eq:H}
\end{equation}

where $J_{12}$ denotes the Mn--Mn exchange interaction, $J_{13}$ and $J_{23}$ represent the Mn--Co exchange interactions, $\mathbf{S}_{1,2}$ are the Mn spin operators, $\mathbf{J}$ is the total angular momentum operator of the Co$^{2+}$ ion, and $D$ is the uniaxial single-ion anisotropy.

After diagonalizing the crystal-electric-field Hamiltonian, the low-energy physics of the Co$^{2+}$ ion is described by the Kramers ground-state doublet,
\begin{equation}
|\psi_{+}\rangle,\qquad
|\psi_{-}\rangle,
\end{equation}

In the basis $\{|\psi_{+}\rangle,|\psi_{-}\rangle\}$, the projected Hamiltonian is

\begin{equation}
H_{\mathrm{eff}}
=
\begin{pmatrix}
\langle\psi_{+}|H|\psi_{+}\rangle &
\langle\psi_{+}|H|\psi_{-}\rangle\\
\langle\psi_{-}|H|\psi_{+}\rangle &
\langle\psi_{-}|H|\psi_{-}\rangle
\end{pmatrix}.
\label{eq:Heff}
\end{equation}

The Mn--Mn exchange interaction does not act on the Co$^{2+}$ degrees of freedom and therefore contributes equally to both states,

\begin{equation}
H_{12}
=
\begin{pmatrix}
J_{12}\mathbf{S}_1\cdot\mathbf{S}_2 & 0\\
0 & J_{12}\mathbf{S}_1\cdot\mathbf{S}_2
\end{pmatrix}.
\end{equation}

Similarly, the projected anisotropy term becomes

\begin{equation}
H_D
=
\begin{pmatrix}
D\langle J_z^2\rangle & 0\\
0 & D\langle J_z^2\rangle
\end{pmatrix},
\end{equation}

where

\begin{equation}
\langle J_z^2\rangle
=
\langle\psi_{+}|J_z^2|\psi_{+}\rangle
=
\langle\psi_{-}|J_z^2|\psi_{-}\rangle .
\end{equation}

Assuming that the Co$^{2+}$ ground doublet is Ising-like,

\begin{equation}
PJ_xP=PJ_yP=0,
\end{equation}

the exchange interaction simplifies to

\begin{equation}
\mathbf{S}\cdot\mathbf{J}
\rightarrow
S_zJ_z,
\end{equation}

such that

\begin{equation}
H_{\mathrm{ex}}
=
J_{13}S_1^zJ_z
+
J_{23}S_2^zJ_z.
\end{equation}

Defining

\begin{equation}
m=
\langle\psi_{+}|J_z|\psi_{+}\rangle,
\end{equation}

time-reversal symmetry requires

\begin{equation}
\langle\psi_{-}|J_z|\psi_{-}\rangle=-m,
\qquad
\langle\psi_{+}|J_z|\psi_{-}\rangle=0.
\end{equation}

Therefore, the projected exchange Hamiltonian is

\begin{equation}
H_{\mathrm{ex}}
=
\begin{pmatrix}
m\left(J_{13}S_1^z+J_{23}S_2^z\right) & 0\\
0 &
-m\left(J_{13}S_1^z+J_{23}S_2^z\right)
\end{pmatrix}.
\end{equation}

Combining all contributions yields

\begin{equation}
H_{\mathrm{eff}}
=
{\small
\begin{pmatrix}
J_{12}\mathbf{S}_1\!\cdot\!\mathbf{S}_2
+
D\langle J_z^2\rangle
+
mX
&
0\\
0&
J_{12}\mathbf{S}_1\!\cdot\!\mathbf{S}_2
+
D\langle J_z^2\rangle
-
mX
\end{pmatrix}
}
\label{eq:Heff_final}
\end{equation}

where

\begin{equation}
X
=
J_{13}S_1^z
+
J_{23}S_2^z.
\end{equation}

The eigenvalues of the projected Hamiltonian are obtained from

\begin{equation}
\det\left(H_{\mathrm{eff}}-EI\right)=0,
\end{equation}

which gives

\begin{equation}
\left(E_0+mX-E\right)
\left(E_0-mX-E\right)=0,
\end{equation}

with

\begin{equation}
E_0
=
J_{12}\mathbf{S}_1\cdot\mathbf{S}_2
+
D\langle J_z^2\rangle.
\end{equation}

Thus, the two eigenvalues are

\begin{equation}
E_1
=
E_0+mX,
\end{equation}

\begin{equation}
E_2
=
E_0-mX.
\end{equation}

The exchange-induced splitting of the ground-state doublet is therefore

\begin{equation}
\Delta
=
|E_1-E_2|
=
2mX,
\end{equation}

or equivalently,

\begin{equation}
\Delta
=
2\langle J_z\rangle
\left(
J_{13}S_1^z
+
J_{23}S_2^z
\right).
\label{eq:gap}
\end{equation}

Now, for chain B, using the Co(2) ground-state wavefunctions given in Eqs.~(8) and (9), we calculate $\langle J_z\rangle$ and obtain,

\begin{equation}
\Delta
=
5.3
\left(
J_{13}S_1^z
+
J_{23}S_2^z
\right).
\end{equation}

from the SpinW modelling, the obtained exchange interactions are
$J_{\mathrm{Mn2-Co2}}$ ($J_{13}$) = 0.8 meV and
$J_{\mathrm{Mn3-Co2}}$ ($J_{23}$) = 0.14 meV.

Therefore,

\[
\Delta
=
5.3 \times 1.5 \times (0.8-0.14)
=
5.2~\mathrm{meV}.
\]

Now, for chain A, for Co(1), using the ground-state wavefunctions given in Eqs.~(6) and (7), we calculated $\langle J_z\rangle$ and obtained

\begin{equation}
\Delta =
5.3\left(
J_{Mn1-Co1}S_1^z
\right)
\end{equation}

From the SpinW modelling, the obtained exchange interaction is
$J_{\mathrm{Mn1-Co1}}$ ($J_{13}$) = 1.1 meV.
Substituting all the values into Equation $30$, we obtain

\[
\Delta
=
5.3\times1.5\times1.1
=
8.7~\mathrm{meV}.
\]

The predicted spin gaps of $5.2$ and $8.7$ meV arise by exchange-induced splitting of the ground-state Kramers doublets of Co(2) and Co(1). These values are consistent with INS-observed spin excitations at about $4$ and $7$~meV, indicating that the two modes possibly originate from the crystallographically different Mn--Co--Mn chains B and A, respectively. Our INS experiments in this energy range have already revealed indications of spin-phonon coupling (see Fig. $3$(e),(f)). Interestingly, previous bulk magnetic studies showed a single-ion relaxation barrier of roughly $40$~K for the Co$^{2+}$ ions \cite{Seikh2017}, which is very near to the energy of the low-lying magnetic excitation around $4$(~$47$K). This correspondence suggests that the Orbach relaxation process is mediated by this low-energy magnetic excitation through spin--phonon coupling, where phonons provide the energy required to thermally populate the excited state and facilitate spin reversal.

From ac susceptibility measurements yielded an activation barrier of approximately $167$~K ($\approx 14.3$~meV) for the single-chain magnet (SCM) relaxation \cite{Seikh2017}. Unlike the single-ion magnet (SIM) barrier, which originates primarily from the strong uniaxial anisotropy of the Co$^{2+}$ ions, the SCM barrier is a collective quantity arising from the combined effects of the Co$^{2+}$ single-ion anisotropy and the intrachain exchange interactions. Interestingly, our INS measurements reveal pronounced spin--phonon coupling over the $5$--$13$~meV energy range (see Fig. $3$(e),(f)), which substantially overlaps with the SCM activation energy. This correspondence suggests that lattice vibrations provide an efficient phonon-assisted relaxation pathway for thermally activated spin reversal across the exchange-enhanced barrier. 

The present results therefore indicate that the exchange-induced spin-gap excitations are strongly coupled to lattice vibrations, and these hybrid spin--phonon modes provide the microscopic pathway for the phonon-assisted relaxation process.

\section{Conclusion}

In this manuscript, we conducted a complete analysis of the one-dimensional single-chain magnet in an unusual inorganic oxide material. The low-energy spin excitation is caused by the two crystallographically different Mn--Co--Mn spin chains, which, despite showing relaxation at low temperatures,  spin excitation persists up to room temperature.  Therefore, even in the absence of typical long-range magnetic order, dynamic magnetic correlations are observed in such single-chain magnets. Machine-learning-based lattice-dynamics models accurately replicate the phonon spectrum and provide support for the experimentally reported spin-phonon interaction, which may assist in the thermally activated relaxation process. Crystal-field analysis was carried out effectively employing experimental and theoretical models, and the ground-state wavefunction was further derived via exchange splitting, which facilitated a fundamental inspection. Thus, the existence of low-energy magnetic excitations provides a plausible microscopic pathway for the thermally activated spin-relaxation process observed in bulk magnetic measurements. These results establish the microscopic origin of the low-energy spin dynamics in Sr$_4$Mn$_2$CoO$_9$ and indicate that its magnetic excitation spectrum is governed by the interaction of crystal-field anisotropy, one-dimensional magnetic exchange, and lattice dynamics. More broadly, this study emphasises the importance of combining neutron spectroscopy with crystal-field, spin-wave, and lattice-dynamics modelling to better understand the microscopic mechanisms governing spin dynamics and relaxation in low-dimensional transition-metal oxides, which will provide useful guidelines for future inorganic single-chain magnet design. The origin of room-temperature spin excitation can be studied further from an application standpoint, additianlly the entire study is relevant from a fundamental physics perspective. The persistence of magnetic excitations up to room temperature further demonstrates the robustness of the underlying spin correlations and motivates future investigations into thermally stable spin dynamics in low-dimensional magnetic materials with potential relevance for spin-based information technologies.

\section{Acknowledgements}

TB greatly acknowledges the Science and Engineering Research Board (SERB) (Project No.: SRG/2022/000044). T.B.\ thanks the Science and Technology Facilities Council (STFC), UK, for providing inelastic neutron scattering beam time on the MERLIN instrument (proposal no.\ RB1820225) and LET (proposal no.\ RB1820275) at the ISIS Neutron and Muon Facility. TB thanks the Oak Ridge National Laboratory (ORNL) for providing access to the INSPIRED software facility for machine-learning-based phonon calculations. DTA would like to thank the EPSRC UK for the funding (Grant No. EP/W00562X/1). WP thank partial support from the Indo-French collaboration, CEFIPRA /IFPACR (7308-1) and the CaeSAR project (ANR-23-EXES-0001). WP acknowledge the LAFICS program, supported by the CNRS. GR acknowledge the Raman-Charpak Fellowship by CEFIPRA.

\section*{Competing interests}
The authors declare no competing financial interests.

\bibliographystyle{apsrev4-2}   
\bibliography{ref}

\end{document}


\title{Supporting Information of "Microscopic investigation of spin dynamics in the single-chain magnet Sr$_4$Mn$_2$CoO$_9$"}

\author{G. Roy}
\email{gourabr22bs@rgipt.ac.in}
\affiliation{Rajiv Gandhi Institute of Petroleum Technology, Jais, Amethi 229304, Uttar Pradesh, India}

\author{S. Ghosh}
\affiliation{Rajiv Gandhi Institute of Petroleum Technology, Jais, Amethi 229304, Uttar Pradesh, India}

\author{M. Kumar}
\affiliation{Rajiv Gandhi Institute of Petroleum Technology, Jais, Amethi 229304, Uttar Pradesh, India}

\author{E. Kushwaha}
\affiliation{Rajiv Gandhi Institute of Petroleum Technology, Jais, Amethi 229304, Uttar Pradesh, India}

\author{J. Sannigrahi}
\affiliation{School of Physical Sciences, Indian Institute of Technology Goa, Ponda 403401, Goa, India}

\author{V. Caignaert}
\affiliation{Laboratoire CRISMAT, Université de Caen Normandie, ENSICAEN, CNRS UMR 6508, Normandie Univ., 14000 Caen, France}

\author{W. Prellier}
\affiliation{Laboratoire CRISMAT, Université de Caen Normandie, ENSICAEN, CNRS UMR 6508, Normandie Univ., 14000 Caen, France}

\author{D. T. Adroja}
\affiliation{ISIS Neutron and Muon Source, STFC, Rutherford Appleton Laboratory, Chilton, Oxon OX11 0QX, United Kingdom}
\affiliation{Highly Correlated Matter Research Group, Physics Department, University of Johannesburg, Auckland Park 2006, South Africa}

\author{D. Voneshen}
\affiliation{ISIS Neutron and Muon Source, STFC, Rutherford Appleton Laboratory, Chilton, Oxon OX11 0QX, United Kingdom}
\affiliation{Department of Physics, Royal Holloway, University of London, TW20 0EX, UK}s

\author{V. Hardy}
\affiliation{Laboratoire CRISMAT, Université de Caen Normandie, ENSICAEN, CNRS UMR 6508, Normandie Univ., 14000 Caen, France}

\author{T. Basu}
\email{tathamay.basu@rgipt.ac.in}
\affiliation{Rajiv Gandhi Institute of Petroleum Technology, Jais, Amethi 229304, Uttar Pradesh, India}

\maketitle

\title{Microscopic investigation of spin dynamics in the single-chain magnet Sr$_4$Mn$_2$CoO$_9$}

\author{G. Roy}
\email{gourabr22bs@rgipt.ac.in}
\affiliation{Rajiv Gandhi Institute of Petroleum Technology, Jais, Amethi 229304, Uttar Pradesh, India}

\author{M. Kumar}
\affiliation{Rajiv Gandhi Institute of Petroleum Technology, Jais, Amethi 229304, Uttar Pradesh, India}

\author{E. Kushwaha}
\affiliation{Rajiv Gandhi Institute of Petroleum Technology, Jais, Amethi 229304, Uttar Pradesh, India}

\author{S. Ghosh}
\affiliation{Rajiv Gandhi Institute of Petroleum Technology, Jais, Amethi 229304, Uttar Pradesh, India}

\author{J. Sannigrahi}
\affiliation{School of Physical Sciences, Indian Institute of Technology Goa, Ponda 403401, Goa, India}

\author{V. Caignaert}
\affiliation{Laboratoire CRISMAT, Université de Caen Normandie, ENSICAEN, CNRS UMR 6508, Normandie Univ., 14000 Caen, France}

\author{W. Prellier}
\affiliation{Laboratoire CRISMAT, Université de Caen Normandie, ENSICAEN, CNRS UMR 6508, Normandie Univ., 14000 Caen, France}

\author{D. T. Adroja}
\affiliation{ISIS Neutron and Muon Source, STFC, Rutherford Appleton Laboratory, Chilton, Oxon OX11 0QX, United Kingdom}
\affiliation{Highly Correlated Matter Research Group, Physics Department, University of Johannesburg, Auckland Park 2006, South Africa}

\author{D. Voneshen}
\affiliation{ISIS Neutron and Muon Source, STFC, Rutherford Appleton Laboratory, Chilton, Oxon OX11 0QX, United Kingdom}

\author{V. Hardy}
\affiliation{Laboratoire CRISMAT, Université de Caen Normandie, ENSICAEN, CNRS UMR 6508, Normandie Univ., 14000 Caen, France}

\author{T. Basu}
\email{tathamay.basu@rgipt.ac.in}
\affiliation{Rajiv Gandhi Institute of Petroleum Technology, Jais, Amethi 229304, Uttar Pradesh, India}

\setcounter{figure}{0}
\renewcommand{\thefigure}{S\arabic{figure}}

\setcounter{table}{0}
\renewcommand{\thetable}{S\arabic{table}}

\begin{figure}[!ht]
    \centering
    \includegraphics[width=.8\textwidth]{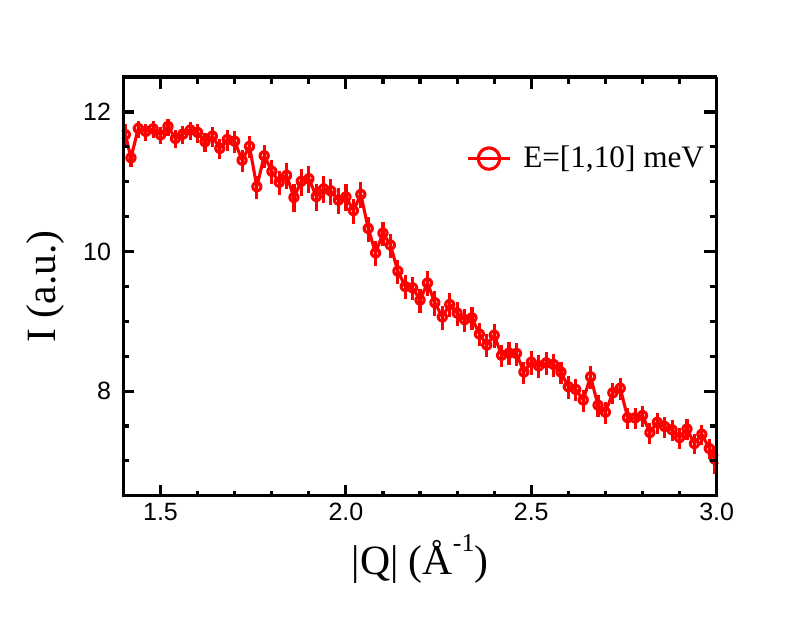}
    \caption{Intensity as a function of momentum transfer ($Q$), integrated over the energy range $E = 1$--$10~\mathrm{meV}$, measured at $2~\mathrm{K}$ using an incident neutron energy of $E_i = 12~\mathrm{meV}$.}
    \label{fig:crystal_structure}
\end{figure}

\begin{figure}[!ht]
    \centering
    \includegraphics[width=1\textwidth]{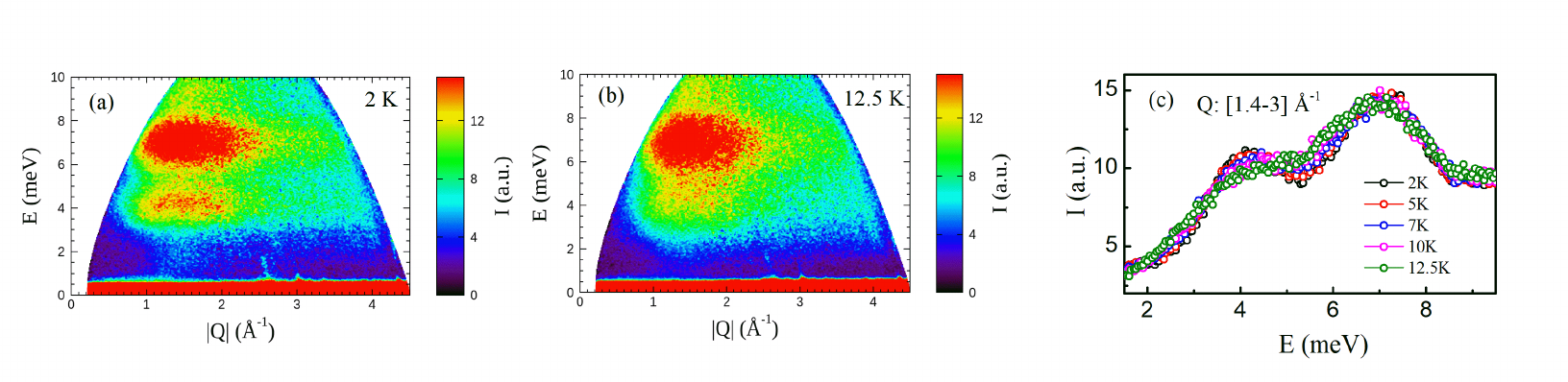}
    \caption{(a,b) Colour contour maps of the Bose-corrected inelastic neutron scattering (INS) intensity measured using the LET spectrometer with an incident energy of $E_i = 12$ meV at 2 K and 12.5 K, respectively. (c) Energy dependence of the Bose-corrected INS intensity integrated over the $Q$ range of $1.4$--$3.0~\mathrm{\AA^{-1}}$ at various temperatures.}
    \label{fig:crystal_structure}
\end{figure}

\begin{figure}[!ht]
    \centering
    \includegraphics[width=0.8\textwidth]{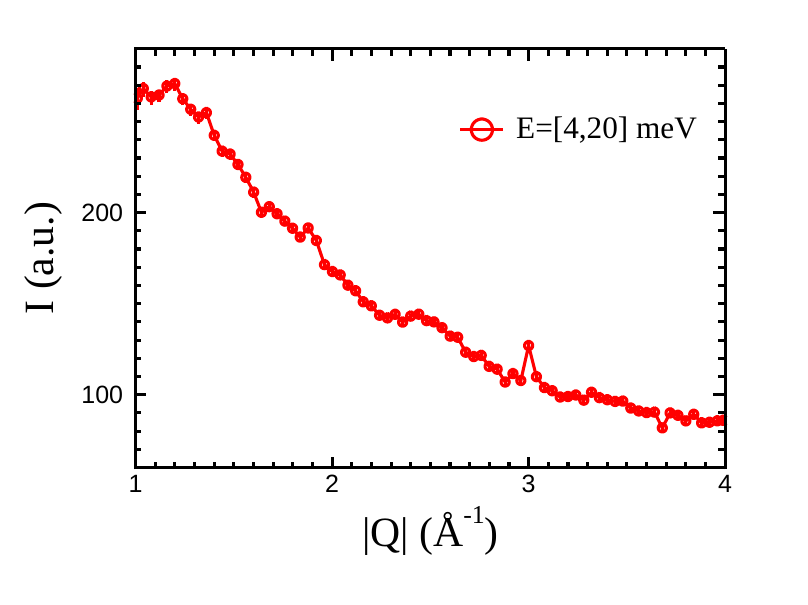}
    \caption{Intensity as a function of momentum transfer ($Q$), integrated over the energy range $E = 5$--$20~\mathrm{meV}$, measured at $7~\mathrm{K}$ using an incident neutron energy of $E_i = 30~\mathrm{meV}$.}
    \label{fig:crystal_structure}
\end{figure}

\begin{figure}[!ht]
    \centering
    \includegraphics[width=0.8\textwidth]{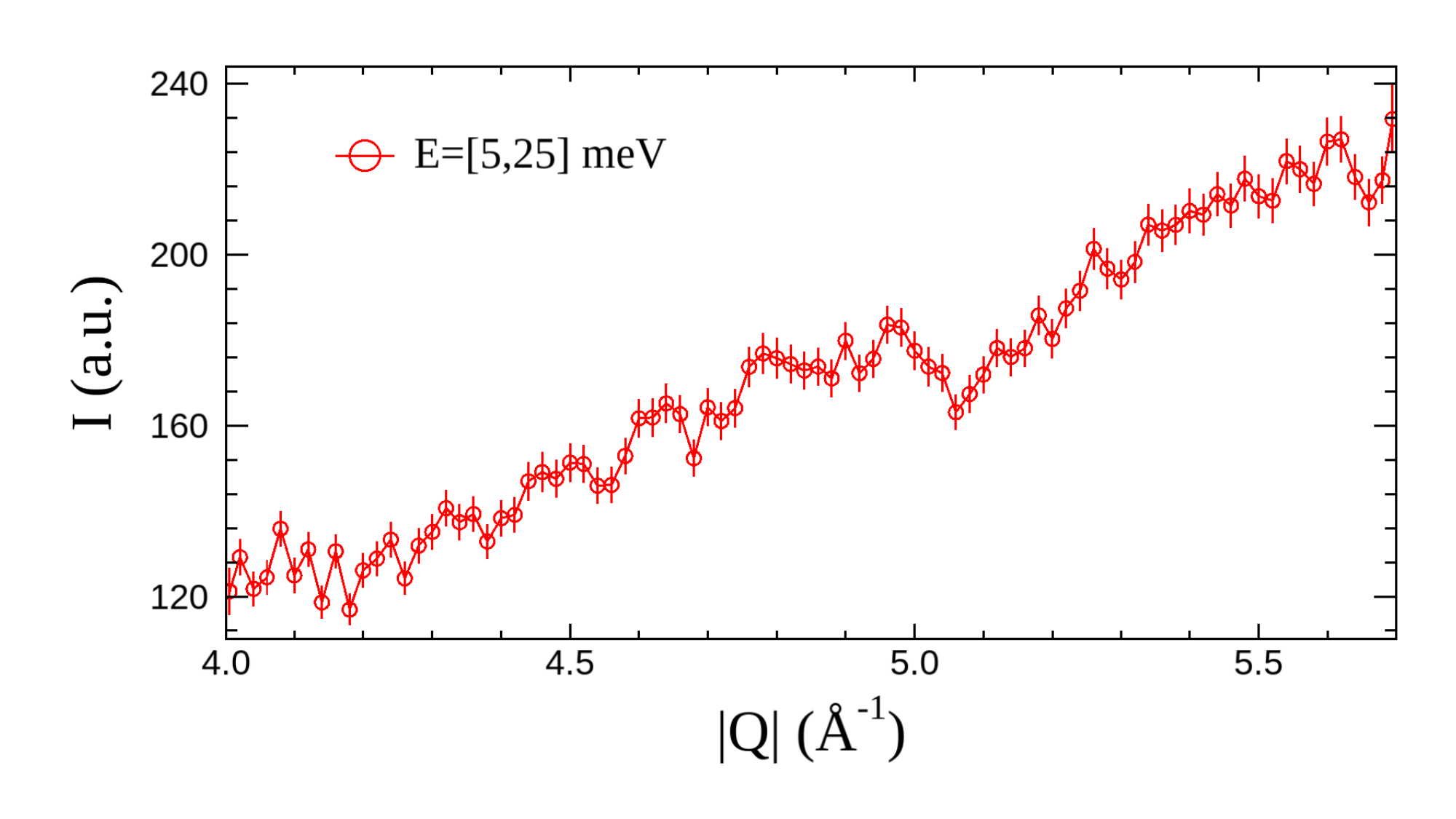}
    \caption{Intensity as a function of momentum transfer ($Q$), integrated over the energy range $E = 5$--$25~\mathrm{meV}$, measured at $300~\mathrm{K}$ using an incident neutron energy of $E_i = 30~\mathrm{meV}$.}
    \label{fig:crystal_structure}
\end{figure}

\begin{figure}[!ht]
    \centering
    \includegraphics[width=1\textwidth]{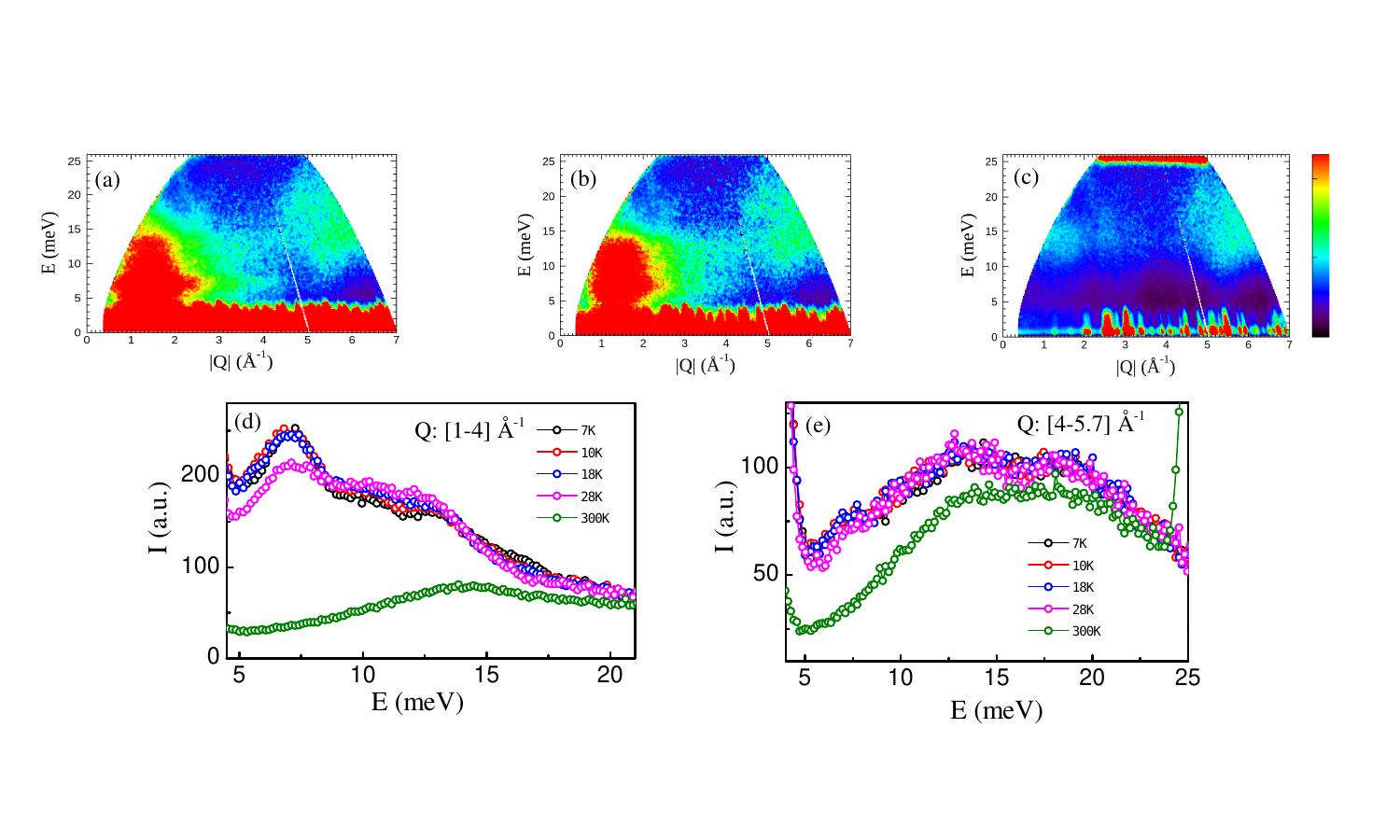}
    \caption{(a--c) Colour contour maps of the  Bose-corrected inelastic neutron scattering (INS) intensity measured using the MERLIN spectrometer with an incident energy of $E_i = 30$ meV at 7 K, 28 K, and 300 K, respectively. (d,e) INS  Bose-corrected intensity as a function of energy transfer, integrated over the $Q$ ranges of $1$--$4$ and $4$--$5.7~\mathrm{\AA^{-1}}$, respectively, at various temperatures.}
    \label{fig:crystal_structure}
\end{figure}

\begin{figure}[!ht]
    \centering
    \includegraphics[width=0.8\textwidth]{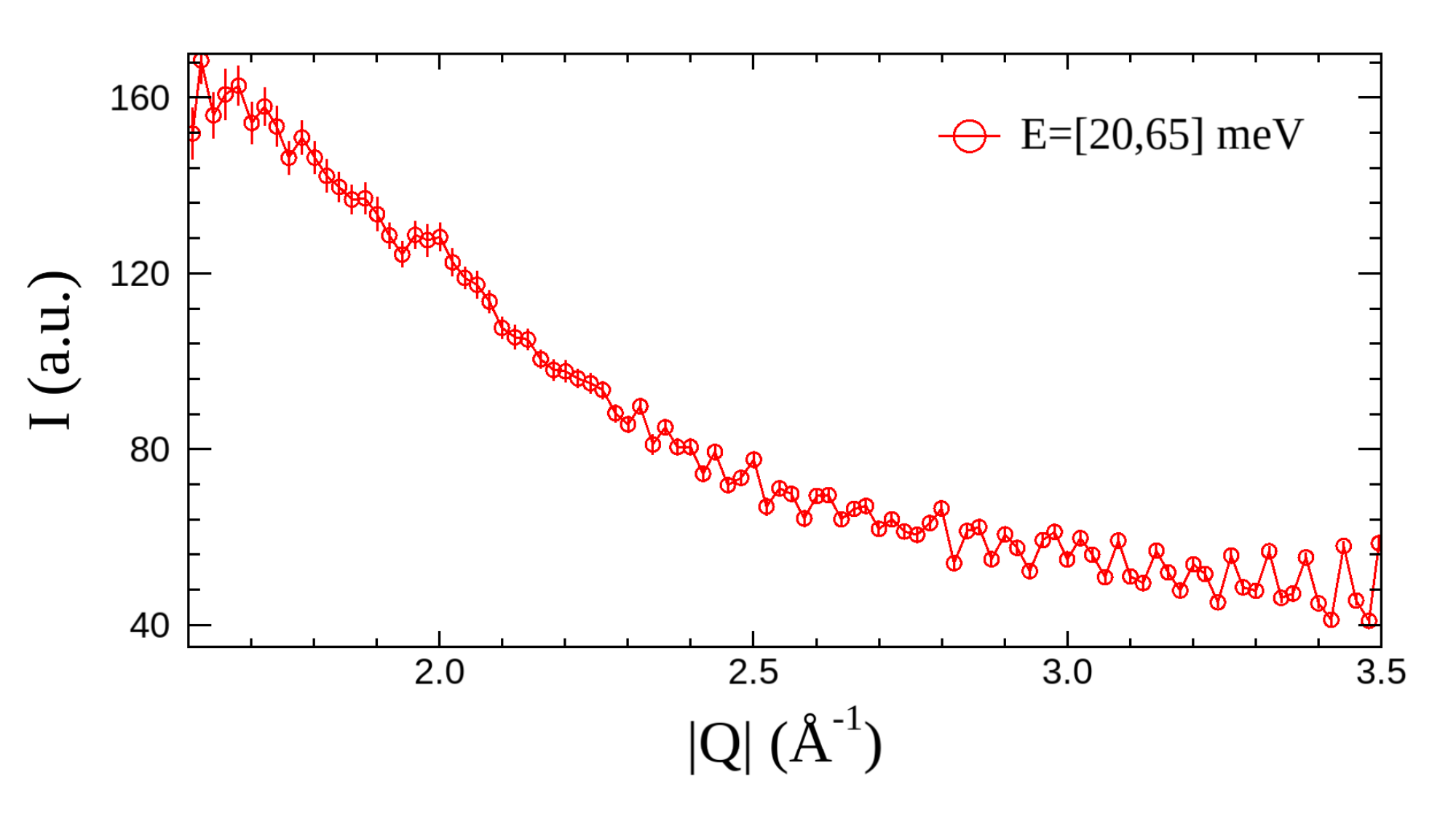}
    \caption{Intensity as a function of momentum transfer ($Q$), integrated over the energy range $E = 20$--$65~\mathrm{meV}$, measured at $7~\mathrm{K}$ using an incident neutron energy of $E_i = 82.1~\mathrm{meV}$.}
    \label{fig:crystal_structure}
\end{figure}

\begin{figure}[!ht]
    \centering
    \includegraphics[width=0.8\textwidth]{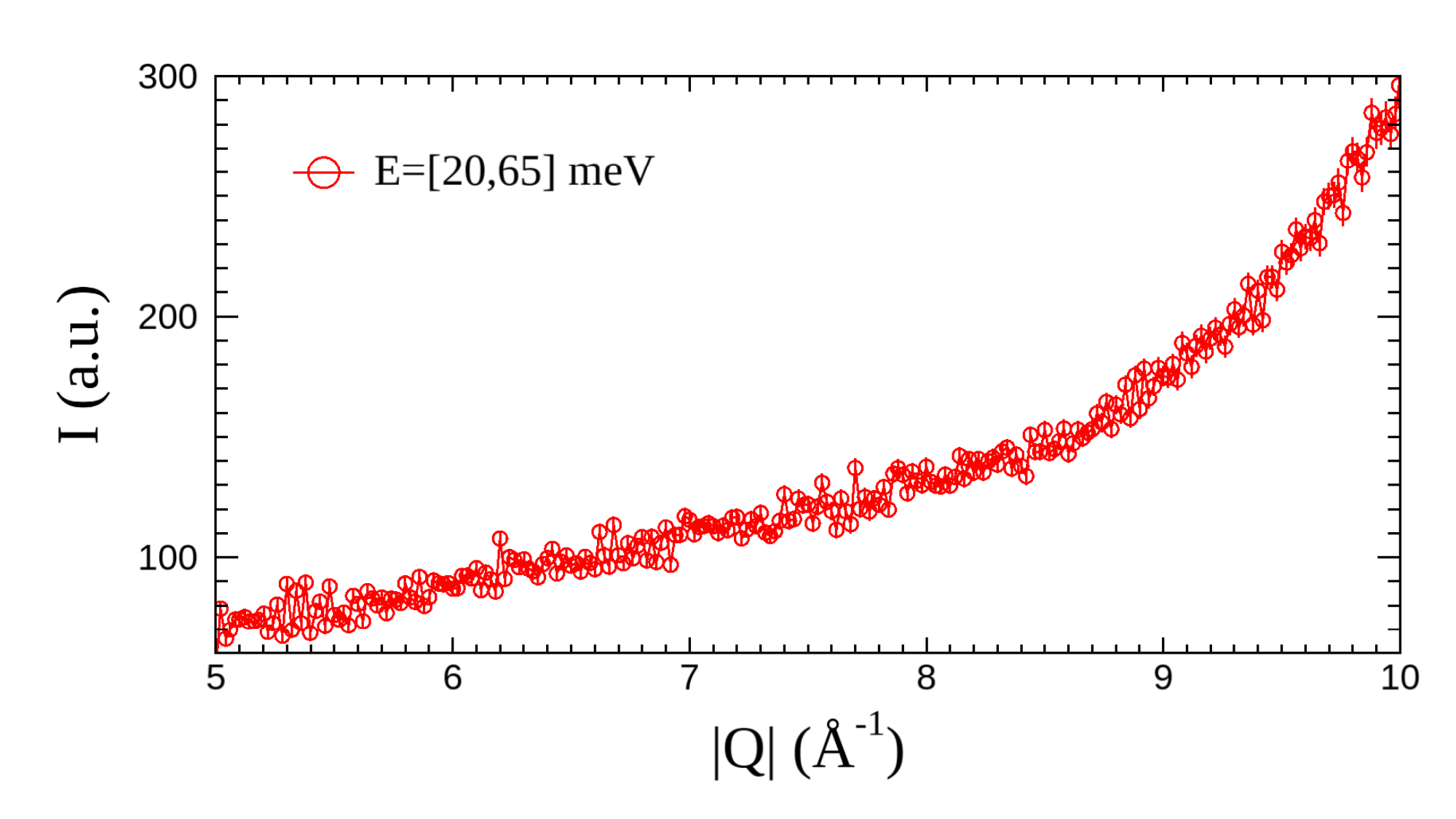}
    \caption{Intensity as a function of momentum transfer ($Q$), integrated over the energy range $E = 20$--$65~\mathrm{meV}$, measured at $300~\mathrm{K}$ using an incident neutron energy of $E_i = 82.1~\mathrm{meV}$.}
    \label{fig:crystal_structure}
\end{figure}